\documentclass[a4paper,preprintnumbers,floatfix,superscriptaddress,prl,twocolumn,showpacs,notitlepage,longbibliography]{revtex4-1}

\usepackage{amsmath, amsthm, amssymb,amsfonts,mathbbol,amstext}
\usepackage{graphicx}
\usepackage{dcolumn}
\usepackage{bm}
\usepackage{bbm}
\usepackage{hyperref}
\usepackage{mathtools}
\usepackage{comment}
\usepackage{color}
\usepackage{multirow}
\usepackage{diagbox}
\usepackage{float}
\usepackage{xcolor}

\def\1{\mathbf{1}}
\def\0{\mathbf{0}}

\DeclareMathOperator{\Tr}{Tr}

\newcommand{\ket}[1]{| #1 \rangle}
\newcommand{\bra}[1]{\langle #1 |}

\newcommand{\processnext}[1]{%
	\ifx\listfinish#1\empty\else\listact{#1}\expandafter\processnext\fi}

\newcommand{\bdm}[1]{d^{\,2M}\!\bm{#1}\,}

\newcommand{\QPD}{\hbox{PQD}}
\newcommand{\sQPD}{\hbox{($s$)-PQD}}
\newcommand{\SQPD}{\hbox{($\bm{S}$)-PQD}}
\newcommand{\SbQPD}{\hbox{($\bar{\bm{S}}$)-PQD}}
\newcommand{\mSQPD}{\hbox{($-\bm{S}$)-PQD}}

\begin{document}
	
	\title{Verification of joint measurability using phase-space quasiprobability distributions}
	
	\author{Saleh Rahimi-Keshari}
	\email{srahimik@ut.ac.ir}
	\affiliation{Department of Physics, University of Tehran, P.O. Box 14395-547, Tehran, Iran}
	\affiliation{School of Nano Science, Institute for Research in Fundamental Sciences (IPM), P.O. Box 19395-5531, Tehran, Iran}
	\author{Mohammad Mehboudi}
	\affiliation{ICFO-Institut de Ciencies Fotoniques, The Barcelona Institute of Science and Technology, 08860 Castelldefels (Barcelona), Spain}
	\affiliation{Max-Planck-Institut f\"ur Quantenoptik, D-85748 Garching, Germany}
	\author{Dario De Santis}
	\affiliation{ICFO-Institut de Ciencies Fotoniques, The Barcelona Institute of
		Science and Technology, 08860 Castelldefels (Barcelona), Spain}
	\author{Daniel Cavalcanti}
	\affiliation{ICFO-Institut de Ciencies Fotoniques, The Barcelona Institute of
		Science and Technology, 08860 Castelldefels (Barcelona), Spain}
	\author{Antonio Ac\'in}
	\affiliation{ICFO-Institut de Ciencies Fotoniques, The Barcelona Institute of
		Science and Technology, 08860 Castelldefels (Barcelona), Spain}
	\affiliation{ICREA, Pg. Lluis Companys 23, 08010 Barcelona, Spain}
	
	\begin{abstract}
		Measurement incompatibility is a distinguishing property of quantum physics and an essential resource for many quantum information processing tasks. We introduce an approach to verify the joint measurability of measurements based on phase-space quasiprobability distributions. Our results therefore establish a connection between two notions of non-classicality, namely the negativity of quasiprobability distributions and measurement incompatibility. We show how our approach can be applied to the study of incompatibility-breaking channels and derive incompatibility-breaking sufficient conditions for bosonic systems and Gaussian channels. In particular, these conditions provide useful tools for investigating the effects of errors and imperfections on the incompatibility of measurements in practice. To illustrate our method, we consider all classes of single-mode Gaussian channels. We show that pure lossy channels with 50\% or more losses break the incompatibility of all measurements that can be represented by non-negative Wigner functions, which includes the set of Gaussian measurements.
	\end{abstract}
	
	\maketitle
	
	A fundamentally distinct feature in quantum mechanics compared to classical physics is the existence of measurements that cannot be performed simultaneously. Examples of such measurements are those corresponding to observables that do not commute, such as position and momentum of a particle \cite{krausbook}. However, commutativity does not entirely capture the notion of measurement incompatibility: for non-projective measurements, described by positive-operator-value measures (POVM), one should employ the notion of joint measurability, defined as follows. A set of $N$ measurements $\{M_x\}_{x=1}^N$, each of them described by measurement operators $M_{a|x}$ for outcomes $a$ such that $M_{a|x}\geq0 ~\forall a,x$ and $\int_a M_{a|x} da=I ~\forall x$ with $I$ being the identity operator, is compatible or jointly measurable if there exists a single measurement $E$ described by measurement operators $\{E_\lambda\}_\lambda$ such that 
	\begin{equation}\label{compatible M}
		M_{a|x}=\int_\lambda \pi(a|x,\lambda) E_\lambda d\lambda,~\forall a,x,
	\end{equation}
	where $\pi(a|x,\lambda)$ is a probability measure \cite{kru}. Otherwise the set of measurements is called incompatible or non-jointly measurable. Equation \eqref{compatible M} implies that all measurements $M_x$ can be implemented by making a single measurement $E$ and classically post-processing the measurement results according to the probability $\pi$. Measurement $E$ is known as the mother measurement.
	
	The incompatibility of quantum measurements seems, at first sight, a limitation. However, with the development of quantum information science it was realised that this phenomenon can be used as a resource for applications such as quantum cryptography \cite{cripto_review}, quantum state discrimination \cite{CHT18,UKSYG18,SSC19} and quantum communication \cite{Guerini19}. Moreover, all the correlations that can be obtained by making compatible measurements on shared entangled multipartite states can be classically simulated \cite{Fine82}. This implies that measurement incompatibility is a requirement to achieve violations of Bell inequalities and also steering~\cite{Quintino2014,Uola2014}. It is therefore a necessary resource for the construction of protocols in the one-sided and fully device-independent scenarios~\cite{NL_review,Steering1,Steering2}. 
		\begin{figure}[b]
		\centering
		\includegraphics[width=0.9\columnwidth]{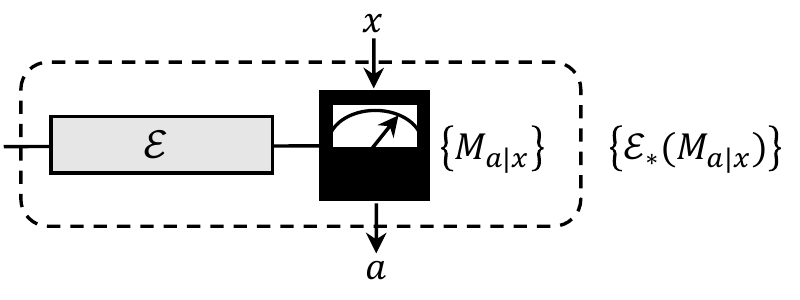}
		\caption{We consider a set of measurements labeled by $x$, with measurement operators $M_{a|x}$ for outcomes $a$, and a quantum channel $\mathcal{E}$. The combination of the channel and measurements can be thought of as a new set of measurements described by $\mathcal{E}_*(M_{a|x})$. Using the phase-space formalism, we investigate the effect of the channel on the incompatibility of measurements.}
		\label{setup}
	\end{figure}

	Given the fundamental and applied importance of measurement incompatibility, it is crucial to derive constructions to identify whether a set of quantum measurements is jointly measurable and, if this is the case, provide a mother POVM. A related question concerns the study of measurement incompatibility under the action of quantum channels. In general, noise-free quantum measurements are incompatible. However, the situation may significantly change in the presence of imperfections. As shown in Fig.~\ref{setup}, suppose that the measurements $\{M_{a|x}\}$ are performed at the output of a fixed quantum channel $\mathcal{E}$. In this case we can consider the combination of the channel and the measurements as a new set of measurements described by measurement operators $\mathcal{E}_{*}(M_{a|x})$. Here, $\mathcal{E}_{*}$ represents the dual channel, defined through $\Tr(\mathcal{E}(\rho)M_{a|x})=\Tr(\rho\mathcal{E}_{*}(M_{a|x}))$. Evidently, incorporating the quantum channel preserves the joint measureability of the measurements, which can be seen using Eq.~(\ref{compatible M}), linearity of the channel, and the fact that $\{\mathcal{E}_{*}(E_\lambda)\}$ defines a valid measurement. However, a quantum channel can have a destructive effect on the incompatibility of measurements, and can make the new set of measurements $\{\mathcal{E}_{*}(M_{a|x})\}$ jointly measurable. Such channels are known as {\it incompatibility breaking channels} \cite{Heinosaari_2015} and their characterization is useful to investigate the effects of noise and errors present in any realistic experiment on quantum information protocols based on measurement incompatibility.
	
	So far, most of the existing works studying these questions have focused on finite dimensional quantum systems~\cite{Heinosaari08,Heinosaari15,SkrCav15,Uola16,Bavaresco17,Designolle2019,Costa18}. Much less is known about the compatibility of measurements on infinite-dimensional continuous-variable (CV) systems, with the exception of the results on particular sets of measurements such as Gaussian measurements or subsets of it~\cite{Heinosaari14,HeinosaariJMP2015,PhysRevA.96.042331}. The question is relevant for a fundamental but also applied point of view, as these measurements are used to describe many relevant quantum setups, e.g. CV quantum optics experiments.

	In this work, we present a general method for studying the joint measurability of a set of measurements based on phase-space quasiprobability distributions (\QPD s) in quantum optics. The method establishes a connection between two notions of nonclassicality: the negativity of the \QPD s representing the measurement operators and the incompatibility of the measurements. We then show how the method provides a practical tool for investigating the effects of noisy channels on the incompatibility of measurements, and use it to derive sufficient conditions for a Gaussian channel to break the incompatibility of different set of measurements, not necessarily Gaussian. For instance, in the case of single-mode loss channels, we show that for losses above or equal to $50\%$, all measurements with non-negative Wigner functions become jointly measurable, extending the previous condition derived only for Gaussian measurements in~\cite{HeinosaariJMP2015}. Our formalism imposes strong limitations on the usefulness of sets of measurements on CV systems for quantum information protocols requiring measurement incompatibility (e.g. one-side and fully device-independent protocols). Moreover, we show that our formalism gives an upper bound on the degree of incompatibility, based on how much noise can destroy the incompatibility of measurements. This bound is tight for Gaussian measurements and Gaussian channels.
		
	
	{\it Phase-space quasiprobability distributions.---}  We start by recalling the phase space formalism, which is at the basis of our results. In this work, we focus on the well-known class of $\bm S$-ordered phase-space quasiprobability distributions [\SQPD{s}] in quantum optics \cite{cahill1969,HILLERY1984,PhysRevX.6.021039}. For the $M$-mode case, they are defined by the family of operators 
	\begin{equation}
		\label{Delta_s}
		\Delta^{(\bm{S})}(\bm z)=\int\frac{\bdm{y}}{(2\pi)^{2M}}\,
		D(\bm y)\, e^{\bm{y}\bm{S}\bm{y}^T/4}\,
		e^{-i\bm{z}\bm{\Omega}\bm{y}^T},
	\end{equation}
	Here $\bm{S}$ is a $2M\times 2M$ symmetric matrix representing the ordering and $D(\bm y)=\exp(-i\bm{y} \bm{\Omega} \bm{X}^T)$ is the displacement (Weyl) operator, where $\bm{X}=(x_1,p_1,\dots, x_M,p_M)$ is the vector of canonical operators $[x_j,p_k]=i\delta_{j,k}$, $\bm{y}, \bm{z} \in \mathbb{R}^{2M}$and $\bm\Omega=\bigoplus_{j=1}^{M}\bigg(\begin{matrix}0 & 1\\ -1 & 0\end{matrix}\bigg)$. Using these operators, as shown in Appendix A, the measurement POVM elements can be written as \cite{cahill1969}
	\begin{equation}
		\label{M_W_rep}
		M_{\bm{a}|\bm{x}}=(2\pi)^M \int \bdm{z} W^{(\bm S)}(\bm{a}|\bm{x},\bm z) \Delta^{(-\bm{S})}(-\bm{z}),
	\end{equation}
	where $W^{(\bm S)}(\bm{a}|\bm{x},\bm z)=\Tr[M_{\bm{a}|\bm{x}}\Delta^{(\bm{S})}(\bm z)]$ is the \SQPD\ representing the measurement operator, and $\bm{x}$ and $\bm{a}$ are, in general, vectors of parameters representing the choices of $M$-mode measurements and their outcomes, respectively. Notice that since $M_{\bm{a}|\bm{x}}$ is Hermitian, $\Delta^{(-\bm{S})}(-\bm{z})$  can be replaced with $\Delta^{(-\bm{S})}(\bm{z})$ in Eq.~(\ref{M_W_rep}). Also, the completeness relation for measurement operators implies $(2\pi)^M\!\int d\bm{a} W^{(\bm S)}(\bm{a}|\bm{x},\bm{z}){=}(2\pi)^M\!\Tr[\Delta^{(\bm{S})}(\bm z)]=1$. 
		
	For a given quantum state $\bm\rho$ one can compute the output probabilities of the measurements using PQDs, 
	\begin{equation}
		\Tr[\rho M_{\bm{a}|\bm{x}}]
		=(2\pi)^M \int \bdm{z} W^{(-\bm S)}(\bm z|\rho) W^{(\bm S)}(\bm{a}|\bm{x},\bm z),
		\nonumber
	\end{equation}
	where $W^{(-\bm S)}(\bm z|\rho)=\Tr[\rho \Delta^{(-\bm{S})}(\bm{z})]$ is \mSQPD\ and  can be viewed as the dual of \SQPD, representing the state $\bm\rho$.  $W^{(-\bm S)}(\bm z|\rho)$ is normalized to one, as  $\int \bdm{z} \Delta^{(-\bm{S})}(\bm{z})=I$. For the special case of $\bm{S}=0$, corresponding to symmetric ordering, the self-dual \QPD\ is the Wigner function. For $\bm{S}=\bm{I}_{2M}$ with  $2M\times 2M$ being identity matrix, PQD becomes the Glauber-Sudarshan $P$ function \cite{Glauber1963,Sudarshan1963}. For $\bm{S}=-\bm{I}_{2M}$ we have the Husimi $Q$-function that is always non-negative for all positive operators~\cite{Husimi}. One can verify that if \hbox{($\bar{\bm{S}}$)-QPD} is non-negative, then all other \hbox{($\bm{S}$)-QPD} with ${\bm{S}}\leq\bar{\bm{S}}$ are given by the convolution of \hbox{($\bar{\bm{S}}$)-QPD} with a Gaussian function, and hence are non-negative as well.
	
	In general, $\bm S$ can be any matrix but if the condition $\bm{S}+i\bm\Omega\geq0$ holds then the operators $\Delta^{(-\bm{S})}(\bm{z})$ are positive and represent the POVM elements of a Gaussian measurement \cite{Kiukas_2013}. To show this, using $D(\bm{y})\exp(-i\bm{z}\bm{\Omega}\bm{y}^T)=D(\bm{z})D(\bm{y})D^\dagger(\bm{z})$ and Eq.~(\ref{Delta_s}), we can write 
		\begin{equation}
			\label{Delta_dis_Gaus}
			\Delta^{(-\bm{S})}(\bm{z})=\frac{1}{(2\pi)^M} D(\bm{z})M_G D^\dagger(\bm{z}),
		\end{equation}
		where $M_G$ is an operator with $\Tr[M_G]=1$ and $\Tr[M_G D(\bm y)]=\exp(-\bm{y}\bm{S}\bm{y}^T/4)$. This relation implies that $\Delta^{(-\bm{S})}(\bm{z})\geq0$ if and only if $M_G\geq0$, which essentially means that $M_G$ must be a Gaussian state with $\bm{S}$ being the covariance matrix of the Wigner function, satisfying the uncertainty relation $\bm{S}+i\bm\Omega\geq0$ \cite{Simon1994,Kiukas_2013}. Notice that if $\Delta^{(-\bm{S})}(\bm{z})$ is positive, $\Delta^{(\bm{S})}(\bm{z})$ cannot be positive as well because $-\bm{S}+i\bm\Omega\geq0$ does not hold. 

		
	{\it Sufficient condition for joint measurability.---} A sufficient conditions for a set of measurements to be jointly measurable follows from the formal analogy between Eqs.~(\ref{compatible M}) and~\eqref{M_W_rep}. If for a set of $N$ measurements $\{M_x\}_{x=1}^N$, there exist positive operators  $\Delta^{(-\bm{S})}(-\bm{z})$ such that $W^{(\bm S)}(\bm{a}|\bm{x},\bm z)\geq0 ~\forall a,x$, then the set is jointly measurable, as $(2\pi)^M W^{(\bm S)}(\bm{a}|\bm{x},\bm z)$ can be viewed as the post-processing of the outcomes of the mother measurement defined by the operators $\Delta^{(-\bm{S})}(-\bm{z})$. We can see that for $\bm S=\bm{I}_{2M}$, $\Delta^{(-\bm{I}_{2M})}(\bm{z})$ is positive and proportional to $M$-mode coherent state. Hence, all measurements with non-negative $P$ functions, known as classical measurements, are jointly measurable. This implies that non-classicality is an essential feature for the incompatibility of measurements.
	
	This approach is particularly useful to study which quantum channels break the incompatibility of a set of measurements. Consider a set of incompatible measurements with non-negative 
	$W^{(\bm S)}(\bm{a}|\bm{x},\bm z)$ that can be expressed in terms of operators $\Delta^{(-\bm{S})}(\bm{z})$, which are not positive. As mentioned above, the effect of a channel $\mathcal{E}$ on these measurements is described by the dual map $\mathcal{E}_*$, getting
	\begin{equation}
		\label{E_class_post-proc}
		\mathcal{E}_*\big(M_{\bm{a}|\bm{x}}\big)=(2\pi)^M\!\!\! \int \bdm{z} W^{(\bm S)}(\bm{a}|\bm{x},\bm z) \mathcal{E}_*\big(\Delta^{(-\bm{S})}(\bm{z})\big).
	\end{equation}
If the channel is such that $\mathcal{E}_*\big(\Delta^{(-\bm{S})}(\bm{z})\big)$ become non-negative bounded operators $\forall\bm{z}$,then it breaks the incompatibility of the measurements in the set. In this case, the operators $\mathcal{E}_*\big(\Delta^{(-\bm{S})}(\bm{z})\big)$, summing up to the identity $\int \bdm{z}\mathcal{E}_*\big(\Delta^{(-\bm{S})}(\bm{z})\big)=\mathcal{E}_*\big(\bm{I}\big)=\bm{I}$, form a POVM for the mother measurement. Notice that if the sufficient condition is not satisfied, it is not guaranteed that the measurements remain incompatible. These sufficient conditions for joint measurability represent our first and most general contribution, which in particular do not require a Gaussian form of neither the measurements nor the channel. 
	
	
	{\it Incompatibility breaking Gaussian channels.---} We illustrate the power of our approach by considering the important case of Gaussian channels, which are readily available in the lab and also used to describe errors in detectors and communication channels. Gaussian channels transform Gaussian states to Gaussian states and are defined by two $2M\times2M$ matrices $N$ and $T$ and a displacement vector $\bm{d}\in\mathbb{R}^{2M}$ \cite{Holevo2001,Weedbrook2012}.  Their action can be fully specified by the application of their dual on the displacement operator
	\begin{equation}
		\label{Gaussian_channel}
		\mathcal{E}_*\big(D(\bm y)\big)=D(\bm{y}\bm{T}) e^{-\bm{y}\bm{N}\bm{y}^T/4-i\bm{d}\bm{\Omega}\bm{y}^T} .
	\end{equation}
	The complete positivity condition of the channel requires $\bm{N}+i\bm{\Omega}-i\bm{T}\bm{\Omega}\bm{T}^T\geq 0$. 
	
	To study these channels, we make use of the well-known \SQPD{s}. By inserting Eq.~(\ref{Gaussian_channel}) into (\ref{Delta_s}), and using the linearity of quantum channels, we find that the action of a Gaussian channel on operators $\Delta^{(-\bm{S})}(\bm{z})$ is
	\begin{equation} 
		\label{E_Gaus_Delta}
		\begin{split}
			\mathcal{E}_*\big(&\Delta^{(-\bm{S})}(\bm{z})\big)=\int\frac{\bdm{y}}{(2\pi)^{2M}}\,
			D\big(\bm{y}\bm{T}\big)\\ &\times\exp\!\left({-\bm{y}(\bm{N}+\bm{S})\bm{y}^T/4-i(\bm{d}+\bm{z})\bm{\Omega}\bm{y}^T}\right).
		\end{split}
	\end{equation}
	For the case of $\bm{S}=0$, this operator is positive definite, corresponding to a Gaussian measurement, if $\bm{N}-i\bm{T}\bm{\Omega}\bm{T}^T\geq 0$~\cite{Kiukas_2013}. Thus, by adding $\bm{S}$ to this condition, we find that $\mathcal{E}_*\big(\Delta^{(-\bm{S})}(\bm{z})\big)$ define an $M$-mode Gaussian measurement if
	\begin{equation}
		\label{Gaus_positive_cond}
		\bm{N}+\bm{S}-i\bm{T}\bm{\Omega}\bm{T}^T\geq 0. 
	\end{equation} 
	This is our second main result, which provides a sufficient condition for incompatibility breaking Gaussian channels. 
	
	Consider a set of incompatible measurements that have non-negative PQDs $W^{(\bar{\bm{S}})}(\bm{a}|\bm{x},\bm z)$ where $\bar{\bm{S}}\leq \bm{I}_{2M}$ is the ordering matrix. The incompatibility of these measurements is broken by any Gaussian channel with matrices $N$ and $T$ that satisfy condition~(\ref{Gaus_positive_cond}). Notice that by finding the maximum ordering matrix $\bar{\bm{S}}$ such that the PQDs are non-negative, we can obtain the minimum $N$ satisfying the condition. The result is constructive: the positive operators $\mathcal{E}_*\big(\Delta^{(-\bm{S})}(\bm{z})\big)$ define the mother measurement, which corresponds to an $M$-mode Gaussian measurement, while the distributions $W^{(\bm S)}(\bm{a}|\bm{x},\bm z)$ specify the post-processing of the measurement outputs, see Eqs.~\eqref{M_W_rep} and~(\ref{E_class_post-proc}).
	Conversely, given a Gaussian channel with $N$ and $T$ matrices, we can see that all measurements whose \SQPD{s} are non-negative for $\bm{S}\geq i\bm{T}\bm{\Omega}\bm{T}^T-\bm{N}$ become jointly measurable under the action of this channel. This result, in particular, shows what measurements should be excluded for steering over noisy channels. In what follows, we focus the analysis on the important class of single-mode Gaussian channels and use our sufficient condition to investigate their effects on the incompatibility of measurements.


	\begin{table*}[t!]
		\begin{tabular}{ |c||c|c|c|p{6cm}|  }
			\hline
			\multicolumn{5}{|c|}{Incompatibility breaking of single mode Gaussian channels}
			\\ \hline
			Class Name & Consideration & The matrix $\bm{T}$ & The matrix $\bm{N}$ & Breaks incompatibility of measurements with positive $W^{(s)}(\bm{a}|\bm{x},\bm z)$ if $s$ satisfies \\
			\hline
			$A_1$  & $\tau = 0$ & $\bm{0}$   & $(2{\bar n} + 1)\bm{I}_2$ & $\forall s$ \\
			$A_2$ & $\tau = 0$ &  $\frac{1}{2}(\bm{Z} + \bm{I}_2)$  & $(2{\bar n} + 1)\bm{I}_2$ &$\forall s$ \\
			$B_1$ & $\tau = 1$ & $\bm{I}_2$ & $\frac{1}{2}(\bm{I}_2 - \bm{Z})$ &  $s\geq s_{\min}= \frac{\sqrt 5 - 1}{2} = 0.618$ \\
			$B_2$ & $\tau = 1$ & $\bm{I}_2$ & ${\bar n} \bm{I}_2$ &  $s\geq s_{\min} = 1-\bar{n}$ \\
			$B_2(Id)$ & $\tau = 1$ & $\bm{I}_2$ & $ \bm{0}$ & $\emptyset$ \\
			${\cal C} ({\rm Loss}) $ & $\tau \in (0,1)$ & $\sqrt{\tau}\bm{I}_2$  & $(1-\tau)(2{\bar n} + 1)\bm{I}_2 $  &$s\geq s_{\min}= \tau (2{\bar n}+2) - (2{\bar n} + 1)$ \\
			${\cal C} ({\rm Amp}) $& $\tau \geq 1$ & $\sqrt{\tau}\bm{I}_2$  & $(\tau-1)(2{\bar n} + 1)\bm{I}_2 $ & $s\geq s_{\min}= 2{\bar n}(1-\tau) + 1 $ \\
			$D$ & $\tau \leq 0$ & $\sqrt{-\tau}\bm{Z}$  & $(1-\tau)(2{\bar n} + 1)\bm{I}_2 $ & $s\geq s_{\min}= 2\tau {\bar n} - (2{\bar n} + 1) $ \\
			\hline
		\end{tabular}
		\caption{Sufficient criteria for incompatibility breaking of single-mode Gaussian channels. Here, $\tau$ is the generalized transmissivity of the channel, $\bar{n}\geq 0$ is the thermal occupation number. The matrix $\bm{Z}$ is the Pauli matrix in the direction of $z$. The trivial channels $A_1$ and $A_2$ break incompatibility of all measurements, while the identity channel $B_2(Id)$ does not affect measurement incompatibility. For all the other channels, our condition \eqref{Gaus_positive_cond} sets a lower bound on $s$ for which all measurements with positive \sQPD\, become compatible. 
		}
		\label{Table_1}
	\end{table*}
	
	{\it Example I: Single-mode pure loss channels.---}Consider 
	first the class of lossy channels, which can be characterized as $\bm{N}=(1-\tau)\bm{I}_2$, $\bm{T}=\sqrt{\tau} \bm{I}_2$ and $\bm{d}=0$ in Eq.~(\ref{Gaussian_channel}), where $\tau$ is the transmissivity of the channel \cite{Weedbrook2012,Holevo2007}. We restrict our analysis to \SQPD\,with $\bm{S}=s\bm{I}_2$, which we denote by \sQPD. Then,  condition (\ref{Gaus_positive_cond}) becomes
	\begin{equation}
		\label{loss_cond}
		s\geq2\tau-1.
	\end{equation}
	According to this condition, a loss channel with transmissivity $\tau$ breaks the incompatibility of all single-mode measurements whose $W^{(2\tau-1)}(a|x,\bm{z})$ are non-negative.
	Here, for lightening the notation, we use $W^{(s)}(a|x,\bm{z})$ instead of $W^{(s\bm{I}_2)}(a|x,\bm{z})$. 
	In this case, using Eq.~(\ref{E_Gaus_Delta}), we can see that the mother measurement is heterodyne, $\mathcal{E}_*\big(\Delta^{(1-2\tau)}(\bm{z})\big)=\ket{\bm{z}}\bra{\bm{z}}/(2\pi)$, where $\ket{\bm{z}}=D(\bm{z})\ket{0}$ is coherent state. The case of loss with excess noise is also discussed in Appendix B.
	
	As an example, let us consider the special class of Gaussian measurements. These measurements have non-negative Wigner function, $W^{(0)}(a|x,\bm{z})\geq0$. Using the condition~(\ref{loss_cond}), we can see that if the transmissivity $\tau\leq 1/2$, or in other words losses are larger than 50\%, all Gaussian measurements become jointly measurable. The results of~\cite{Heinosaari14,PhysRevA.96.042331} imply that this is a necessary and sufficient condition. In fact, our result is more general, as it applies to all measurements with non-negative Wigner function, a set that strictly includes the set of Gaussian measurements. This condition provides a sort of analog of the detection loophole: when losses are larger than 50\%, no quantum state can violate a steering or Bell inequality using measurements with non-negative Wigner function, such as Gaussian measurements.
	
	It is interesting to note that for measurements whose \sQPD s are non-negative for $s\leq -1$ only, such as photon-counting or photo-detection measurements, condition~\eqref{Gaus_positive_cond} is not satisfied for any transmissivity $0<\tau\leq1$. Nonetheless, in a more realistic scenario, one has to include random counts arising from dark counts, mode mismatching, and other sources of noise that affect the measurement \cite{PhysRevX.6.021039,BARNETT199845}. Denoting the probability of the random counts with $P_D$ we can describe the POVM elements of realistic photo-detection (rpd) with $M_{0|{\rm rpd}} = (1-P_D) \ket{0}\bra{0}$, and $M_{\bar{0}|{\rm rpd}} = I - M_{0|{\rm rpd}}$, which reduces to the ideal photo-detection if $P_D=0$. The corresponding \sQPD s read [see the Appendix C for details]
	\begin{align}\label{eq:realistic_photo_det_SQPD}
		\begin{split}
		W^{(s)}(0|{\rm rpd},\bm{z}) &= \frac{1-P_D}{\pi (1-s)}e^{-|\bm{z}|^2/(1-s)},\\
		W^{(s)}(\bar{0}|{\rm rpd},\bm{z}) & = \frac{1}{2\pi}-W^{(s)}(0|{\rm rpd},\bm{z}).	
		\end{split}
	\end{align}
	These \sQPD s are both positive for $s \leq 1- 2(1-P_D)$. Comparing with \eqref{loss_cond} we conclude that the realistic photo-detection becomes reproducible by heterodyne detection and classical postprocessing if $\tau \leq 1-(1-P_D)$. Moreover, this measurement is compatible with all measurements with non-negative Wigner function for $\tau \leq \min \{1/2 ,  ~ 1-(1-P_D)\}$.
	
	{\it Example II: General single-mode Gaussian channels.---}Single 
	mode Gaussian channels can be classified into eight major groups---up to Gaussian unitaries that will not affect measurement incompatibility---depending on the matrices $\{\bm{N},\bm{T}\}$ which characterizes them \cite{Weedbrook2012}. By choosing $\bm{S}=s\bm{I}_2$, the condition \eqref{Gaus_positive_cond} sets a sufficient criterion for each of these channels to break incompatibility of measurements whose \sQPD, $W^{(s)}(\bm{a}|\bm{x},\bm z)$, is non-negative. We have summarized these criteria in Table~\ref{Table_1}.
	
	{\it Degree of incompatibility.---} One can think of measures of incompatibility in terms of the amount of noise that make a set of measurements jointly measurable~\cite{Heinosaari15,SkrCav15,Uola16,Bavaresco17,Designolle2019}. To define such a measure one would need a sufficient and necessary condition for the incompatibility breaking of a given channel. Our formalism, in general, can provide an upper bound on the degree of incompatibility of a set of measurements. However, for a Gaussian channel and a set of Gaussian measurements this bound can be tight~\cite{HeinosaariJMP2015}. For a set of single-mode Gaussian measurements, the maximum value of the ordering parameter $\bar{s}$ such that $W^{(\bar{s})}(a|x,\bm{z})$ are Gaussian functions is $0\leq \bar{s}\leq 1$. Considering a pure loss channel, as an example, and using Eq.~(\ref{loss_cond}), we can use the maximum transmissivity $\bar{\tau}=(\bar{s}+1)/2$ for incompatibility breaking, to define $d=1-\bar{\tau}=(1-\bar{s})/2$ as a measure of incompatibility. For homodyne measurements we have $d=1/2$, for heterodyne and other classical measurements $d=0$, and for measurements in the displaced-squeezed vacuum basis $0<d<1/2$.
	
	
	{\it Discussion.---}In this work, we have established a connection between the negativity of phase-space quasi-probability distributions and the joint measurability of quantum measurements, both known as useful resources in quantum information processing. This connection provides a new insight into the problem of joint measurability and enables us to formalize sufficient conditions for investigating the effect of quantum channels on the incompatibility of measurements. Our results are constructive, in the sense that they specify a mother measurement and post-processing for the compatible measurements. The derived conditions also provide noise thresholds that need to be satisfied for the observation of Bell or steering inequality violations using relevant sets measurements.
	
	As discussed, the Husimi $Q$ function is non-negative for all measurement operators, so if $\mathcal{E}_*\big(\Delta^{(\bm{I}_{2M})}(\bm{z})\big)\geq0$ the channel breaks the incompatibility of all measurements. But we know that Gaussian channels satisfying this condition, i.e., $\bm{N}-\bm{I}_{2M}-i\bm{T}\bm{\Omega}\bm{T}^T\geq 0$, are also entanglement-breaking channels~\cite{Holevo2008entanglement}. An interesting question is whether there exist quantum channels that are not entanglement breaking but break the incompatibility of all measurements.
	
	Our formalism can be generalized in terms of other quasiprobability distributions, in particular, for finite-dimensional systems~\cite{Ferrie_2009,Ferrie_2011}. In the general context, quasiprobability distributions are associated with pairs of dual frames, $\{G(\lambda)\}$ and $\{F(\lambda)\}$ that we can assume to be normalized: $\int d\lambda G(\lambda)= I$ and $\Tr[F(\lambda)]=1$. Measurement operators can be expressed as
		\begin{equation}
			\label{M_frame}
			M_{a|x}=\int d\lambda V(a|x,\lambda) G(\lambda),
		\end{equation} 
		where $V(a|x,\lambda)=\Tr[M_{a|x} F(\lambda)]$ ($\int da V(a|x,\lambda)=1$) is a quasiprobability representation of the measurement operator. Following the same arguments discussed in the paper, if $G(\lambda)$ are positive, a set of measurements whose $V(a|x,\lambda)\geq0 ~\forall a,x$ are jointly measurable. Likewise, these quasiprobability distributions can be used to verify incompatibility breaking channels. The generalization of our formalism and its applications in quantum protocols deserve further investigation and we leave it as a subject for future research.
	
	
	The authors would like to acknowledge constructive discussions with T. Heinosaari, R. Uola and T. Osborne. This work was financially supported by the Government of Spain (FIS2020-TRANQI and Severo Ochoa CEX2019-000910-S), Fundacio Cellex and Fundacio Mir-Puig, Generalitat de Catalunya (SGR 1381, QuantumCAT and CERCA Programme), the EU project CiviQ, the ERC AdG CERQUTE and the AXA Chair in Quantum Information Science. DC acknowledges a Ramon y Cajal fellowship.
	\bibliographystyle{apsrev4-1}
	\bibliography{Refs}
	\newpage
	\onecolumngrid
	\appendix

\section{Appendix A: The phase space formalism}
In the main text, the operators $\Delta^{(\bm{S})}(\bm z)$ are used to define the $\bm{S}$-ordered phase-space quasiprobability distributions [\SQPD s], $W^{(\bm S)}(O,\bm z)=\Tr[O\Delta^{(\bm{S})}(\bm z)]$, corresponding to the operator $O$ that can be an observable or a density operator. For a given quantum state $\rho$ and any set of POVM operators $\{M_{{\bm{a}|\bm{x}}}\}$---with ${\bm x}$ labeling the specific choice of POVM and ${\bm a}$ the different outcomes for that given choice---we are interested in the outcome probabilities given by the Born rule,
\begin{equation}
	P({\bm{a}|\bm{x}}) = \Tr[\rho M_{\bm{a}|\bm{x}}].\label{eq:Born_rule}
\end{equation}
To show how any $M$-mode operator like $\rho$ can be represented using \SQPD s, we start by expanding the operator in terms of displacement operators~\cite{cahill1969}
\begin{align}
	\begin{split}
\rho & = \frac{1}{(2\pi)^{M}}\int d^{2M}\bm{y} \Tr[\rho D(\bm{y})] D(-\bm{y})\\
& = \frac{1}{(2\pi)^{M}}\int d^{2M}\bm{y} \Tr[\rho D(\bm{y})] D(-\bm{y}) e^{\bm{y}\bm{S}\bm{y}/4} e^{-\bm{y}\bm{S}\bm{y}/4},
	\end{split}\label{eq:spqd_1}
\end{align}
where $\bm{y} \in \mathbb{R}^{2M}$ and in the second line we multiplied $\exp\!\big(\bm{y}\bm{S}\bm{y}/4\big)$ and its inverse with $\bm{S}$ being a $2M\times 2M$ symmetric matrix that can be associated with the ordering of displacement operators. By definition the \mSQPD s for the density operator $\rho$ is given by
\begin{align}
	W^{(-\bm{S})}(\bm{z}|\rho) & = \Tr\big[\rho\Delta^{(-\bm{S})}(\bm z)\big] = \int\frac{\bdm{y}}{(2\pi)^{2M}}\,
	{\rm Tr}[\rho D(\bm y)]\, e^{-\bm{y}\bm{S}\bm{y}^T/4}\,
	e^{-i\bm{z}\bm{\Omega}\bm{y}^T},
	\label{WS}
\end{align}
Here, the operators $\Delta^{(\bm{S})}(\bm z)$ are defined as
\begin{equation}
\Delta^{(\bm{S})}(\bm z)= \int\frac{\bdm{y}}{(2\pi)^{2M}}\,
 D(\bm y)\, e^{\bm{y}\bm{S}\bm{y}^T/4}\,
e^{-i\bm{z}\bm{\Omega}\bm{y}^T},
\end{equation} 
and satisfy the following relations
\begin{equation}
\Tr[\Delta^{(\bm{S})}(\bm z)]=\int\frac{\bdm{y}}{(2\pi)^{2M}}\,
\Tr[D(\bm y)]\, e^{\bm{y}\bm{S}\bm{y}^T/4}\,
e^{-i\bm{z}\bm{\Omega}\bm{y}^T}=\frac{1}{(2\pi)^M}
\label{PQD-identity}
\end{equation}
since $\Tr[D(\bm y)]=(2\pi)^M\delta^{2M}(\bm{y})$, and 
\begin{equation}
\int d^{2M}\bm{z}\, \Delta^{(\bm{S})}(\bm z)= \int\frac{\bdm{y}}{(2\pi)^{2M}}\,
D(\bm y)\, e^{\bm{y}\bm{S}\bm{y}^T/4}\,
\int d^{2M} \bm{z} e^{-i\bm{z}\bm{\Omega}\bm{y}^T}=\bm{I},
\end{equation}
where we used 
\begin{align}
	\int d^{2M} \bm{z} e^{-i\bm{z}\bm{\Omega}\bm{y}^T} = (2\pi)^{2M} \delta^{2M}(\bm{y}).\label{Four-delta-func}
\end{align}

By taking the inverse Fourier transform of Eq.~(\ref{WS}), using Eq.~(\ref{Four-delta-func}), one obtains 
\begin{align}
	\int d^{2M}\bm{z}\, W^{(-\bm{S})}(\bm{z}|\rho) e^{i\bm{z}\bm{\Omega}\bm{y}^T}=
	\Tr[\rho D(\bm{y})]e^{-\bm{y}\bm{S}\bm{y}/4}. \label{eq:spqd_2}
\end{align}
Substituting \eqref{eq:spqd_2} into \eqref{eq:spqd_1} gives
\begin{align}
	\begin{split}
	\rho & = \frac{1}{(2\pi)^{M}}\int d^{2M} \bm{z}\, W^{(-\bm{S})}(\bm{z}|\rho) \int d^{2M}\bm{y}  D(-\bm{y})\,e^{\bm{y}\bm{S}\bm{y}/4} e^{i\bm{y}\Omega \bm{y}^T} \\
	&= (2\pi)^{2M}\int d^{2M} \bm{z}\, W^{(-\bm{S})}(\bm{z}|\rho) \Delta^{(\bm{S})}(\bm{z}).
	\end{split}
\end{align}
Finally, if we replace this in Eq.~\eqref{eq:Born_rule} we obtain
\begin{equation}
	P({\bm{a}|\bm{x}})=\Tr[\rho M_{\bm{a}|\bm{x}}]
	=(2\pi)^M \int \bdm{z} W^{(-\bm S)}(\bm z|\rho)\, W^{(\bm S)}(\bm{a}|\bm{x},\bm z),
\end{equation}
where \SQPD s, $W^{(\bm S)}(\bm{a}|\bm{x},\bm z)=\Tr\big[M_{\bm{a}|\bm{x}} \Delta^{(\bm{S})}(\bm{z})\big]$, represent the measurement operators. 

Notice that the relation between two $\bm{S}$-ordered and $\bar{\bm{S}}$-ordered \QPD s, if $\bar{\bm{S}}-\bm{S}\geq0$ can be understood in terms of the convolution, 
\begin{equation}
\Delta^{(\bm{S})}(\bm z)=\int\! \bdm{k} \frac{\exp({-\bm{k}(\bar{\bm{S}}-\bm{S})^{-1}\bm{k}^T})}{\pi^M \sqrt{\det(\bar{\bm{S}}-\bm{S})}}  \Delta^{(\bar{\bm{S}})}(\bm{z}-\bm{k}).
\end{equation}  
This implies that {\SQPD} can written as the convolution of \SbQPD\ with a Gaussian function, and hence if \SbQPD\ is non-negative all other {\SQPD}s with  ${\bm{S}}\leq\bar{\bm{S}}$ are non-negative as well.

	\section{Appendix B: Single mode loss channel with excess noise}
	A more realistic channel compared to pure lossy channels may also contain some excess noise form the environment. These channels are characterized by 
	\begin{align}\label{Channel_Loss_Excess}
		\bm{N}=(1-\tau+2\epsilon)\bm{I}_2, \hspace{1cm} \bm{T}=\sqrt{\tau} \bm{I}_2,
	\end{align}
	with $\epsilon\geq 0$ quantifying the excess noise. The condition (\ref{Gaus_positive_cond}) now reads
	\begin{equation}
		\label{loss_excess_cond}
		s\geq2\tau-2\epsilon-1, \text{ or } \tau \leq \frac{s+2\epsilon+1}{2}.
	\end{equation}
	Therefore, this channel breaks the incompatibility of all measurements with a non-negative $W^{(2\tau - 2\epsilon - 1)}(a|x,z)$. As a special case, if $\epsilon = \tau$, the incompatibility of all measurements is broken---since the Q-function is always positive. Nonetheless, for $\epsilon = \tau$ one can check that the condition $\bm{N}-\bm{I}_{2M}-i\bm{T}\bm{\Omega}\bm{T}^T=	\tau\big(\bm{I}_2-i \Omega\big) \geq 0$ is satisfied, implying that the channel is also entanglement breaking~\cite{Holevo2008entanglement}. To verify this condition, we can also check when entanglement of two-mode squeezed vacuum states breaks under the action of such channel, by using the entanglement criterion for Gaussian quantum systems in \cite{PhysRevLett.84.2726}. 
	Recall that the covariance matrix of an $M$ mode system with density matrix $\rho$ contains all second order moments, that is $\sigma_{ij} \equiv {\rm Tr}[\rho \{\bm{X}_i~,~\bm{X}_j\}_+]-2{\rm Tr}[\rho \bm{X}_i] {\rm Tr}[\rho \bm{X}_j]$ where $\bm{X} = (x_1,p_1,\dots,x_M,p_M)$ and $\{~,~\}_+$ is the anti-commutator. In particular, a two-mode covariance matrix $\sigma_{AB}$ represents a separable state if and only if $L\sigma_{AB}L + \Omega \geq 0$ with $L = {\rm diag}[1 ~~1~~ 1~-1]$. 
In our case, the covariance matrix of any two mode Gaussian state after the application of the channel on one of the modes is transferred as follows
	\begin{align}
		\sigma_{AB} \to \sigma_{AB}({\tau,\epsilon}) = (\bm{T}\oplus \bm{I}_2)~\sigma_{AB}~(\bm{T}^T\oplus \bm{I}_2) + \bm{N}\oplus \bm{0_2},
	\end{align}
	where $\bm{0}_2$ is the null matrix of dimension two. 
	Let our input into the channel to be the 2-mode squeezed state with the covariance matrix 
	\begin{align}\label{eq:2-mode-squeezed-CV}
		\sigma_{AB} = \left[
		\begin{array}{cc}
			\nu \bm{I}_2 & \sqrt{\nu^2 - 1} \bm{Z}\\
			\sqrt{\nu^2 - 1} \bm{Z} & \nu \bm{I}_2
		\end{array}
		\right],
	\end{align}
	where $\nu\geq 1$ is the squeezing parameter.
	The output state reads
	\begin{align}\label{eq:2-mode-squeezed-CV_double_channel}
		\sigma_{AB}(\tau,\epsilon) = \left[
		\begin{array}{cc}
			K \bm{I}_2 & \sqrt{\tau(\nu^2 - 1)} \bm{Z}\\
			\sqrt{\tau(\nu^2 - 1)} \bm{Z} & \nu \bm{I}_2
		\end{array}
		\right],
	\end{align}
	with $K = 1+2\epsilon + \tau (\nu - 1)$. One can verify that for $\eta\leq\epsilon$ the smallest eigenvalue of $L\sigma_{AB}(\eta,\epsilon)L + \Omega$ is always positive, thus the channel is entanglement breaking. Notice that separable states cannot be used for quantum steering, which implies that all local measurements on the party where the channel is applied become compatible. 
	\section{Appendix C: Examples of non-Gaussian measurements}
		\subsection{Ideal and realistic photo-detection}
	The ideal photo-detection (ipd) measurment can be described by two POVM elements corresponding to no-click or click:
	\begin{align}
		M_{0|{\rm ipd}} = \ket{0}\bra{0}, ~~M_{\bar{0}|{\rm ipd}} = \bm{I} - M_{0|{\rm ipd}},
	\end{align}
	respectively, where $\bm{I}$ is the identity operator. The \sQPD s of the POVM element $M_{0|{\rm ipd}}$ is given by
		\begin{align}
		\begin{split}
			W^{(s)}(0|{\rm ipd},\bm{z}) &=\Tr[M_{\bar{0}|{\rm ipd}} \Delta^{(s)}(\bm{z})]
			=\frac{1}{(2\pi)^{2}} \int d^2\bm{y} \Tr[M_{0|{\rm ipd}} D(\bm{y})] e^{s|\bm{y}|^2/4} e^{-i\bm{z} \Omega \bm{y}^T}\\
			&=\frac{1}{(2\pi)^{2}} \int dy_1 dy_2 e^{-(y_1^2+y_2^2)/4} e^{s(y_1^2+y_2^2)^2/4} e^{iy_1 z_2-iy_2 z_1}\\
			&=\frac{1}{(2\pi)^{2}} \int dy_1 e^{-(1-s)y_1^2/4+i y_1 z_2} \int dy_2 e^{-(1-s)y_2^2/4-i y_2 z_1}\\
			&=\frac{1}{\pi(1-s)}e^{(-z_1^2-z_2^2)/(1-s)}=\frac{e^{{-|\bm{z}|^2}/({1-s})}}{\pi (1-s)}.
		\end{split}
	\end{align}
	The \sQPD s of the second POVM element can be obtained as
	\begin{equation}
		\begin{split}
	W^{(s)}(\bar{0}|{\rm ipd},\bm{z}) & =\Tr[\Delta^{(s)}(\bm{z})]-\Tr[M_{\bar{0}|{\rm ipd}} \Delta^{(s)}(\bm{z})]\\
	&=\frac{1}{2\pi}-W^{(s)}(0|{\rm ipd},\bm{z}),
		\end{split}
	\end{equation}
	where in the second line Eq.~(\ref{PQD-identity}) is used. Notice that $	W^{(s)}(0|{\rm ipd},\bm{z})$ is always positive, but $W^{(s)}(\bar{0}|{\rm ipd},\bm{z})$ has negativity except for the trivial cases $s\leq -1$. Therefore, under the pure loss channel our sufficient condition of incompatibility breaking for \textit{ideal} photo-detection and other measurements is not satisfied.
	
	For the realistic photo-detection scenario including the random counts, the POVM elements are  $M_{0|{\rm rpd}} = (1-P_D) M_{0|{\rm ipd}}$, and $M_{\bar{0}|{\rm rpd}} = \bm{I} - M_{0|{\rm rpd}}$. The corresponding \sQPD s are obtained trivially from the ideal photo-detection and are presented in Eqs.~\eqref{eq:realistic_photo_det_SQPD}.
	\subsection{Thermal photo detection}
	Here we introduce the thermal photo-detection as another example of non-Gaussian measurements that can become compatible with all Gaussian measurements under the pure loss channel. The first POVM element, $M_{T|{\rm tpd}} = e^{-H/T}/Z$, is a thermal state with $H$ being the Hamiltonian and $Z=\Tr[e^{-H/T}]$, 
	 and the other POVM element is $M_{\bar{T}|{\rm tpd}} = \bm{I} - M_{T|{\rm tpd}}$. The characteristic function of a thermal state, which is a Gaussian state, can be found using its covariance matrix $\sigma_T = \coth(1/2T) {\bm I}_2$ and reads
	\begin{align}
		\Tr[D(\bm{y}) M_{T|{\rm tpd}}] = e^{-\nu|\bm{y}|^2/4},
	\end{align}
	with $\nu=\coth(1/2T)\geq 1$. Notice that at zero temperature this measurement is equivalent to the ideal photo-detection.
	The \sQPD\ of $M_{T|{\rm tpd}}$ is given by
	\begin{align}\label{eq:sQPD_thermal_photo_detection}
		\begin{split}
W^{(s)}(T|{\rm tpd},\bm{z}) &= \Tr[M_{T|{\rm tpd}} \Delta^{(s)}(\bm{z})]=\frac{1}{(2\pi)^{2}} \int d^2\bm{y} \Tr[M_{T|{\rm tpd}} D(\bm{y})] e^{s|\bm{y}|^2/4} e^{-i\bm{z} \Omega \bm{y}^T}\\
&=\frac{1}{(2\pi)^{2}} \int dy_1 dy_2 e^{-\nu(y_1^2+y_2^2)/4} e^{s(y_1^2+y_2^2)^2/4} e^{iy_1 z_2-iy_2 z_1}\\
&=\frac{1}{\pi(\nu-s)}e^{(-z_1^2-z_2^2)/(\nu-s)}=\frac{e^{{-|\bm{z}|^2}/({\nu-s})}}{\pi (\nu-s)},
		\end{split}
	\end{align}
	which is always positive. However, the non-negativity of $W^{(s)}(\bar{T}|{\rm tpd},\beta)=1/(2\pi)-W^{(s)}(T|{\rm tpd},\bm{z})$ requires 
	\begin{align}
		\frac{2}{\nu-s}\leq 1 \rightarrow s \leq \nu-2.
	\end{align}
	Firstly, notice that, for $T=0$, we have $\nu=1$, hence the criterion is satisfied only if $s \leq -1$. This is indeed what we found for ideal photo-detection.
	For any other temperature, however, there exist $s > -1$ such that the \sQPD\ is non-negative. To sum up, under the Gaussian channel \eqref{Channel_Loss_Excess} this \textit{non-Gaussian} measurement becomes compatible with Gaussian measurements and all measurements with non-negative Wigner functions if $\nu \geq 2$ and $\tau\leq({1+2\epsilon})/2$. More generally, this measurement becomes compatible with all Gaussian measurements for $\tau \leq \min\{\frac{1+2\epsilon}{2}, \frac{\nu-1+2\epsilon}{2}\}$.

\end{document}